\documentclass[apj,numberedappendix,epsfigure]{emulateapj}

\usepackage{apjfonts}

\shorttitle{ECLIPSING DOUBLE WD BINARY NLTT 11748}
\shortauthors{STEINFADT ET AL.}

\newcommand{\be}{\begin{eqnarray}}
\newcommand{\ee}{\end{eqnarray}}

\newcommand{\kms}{{\rm km\,s}^{-1}}
\newcommand{\mc}{\multicolumn}

\newcommand{\phz}{\phantom{0}}

\begin{document}

% -----------------------------------------------------------
% -----------------------------------------------------------

\title{Discovery of the Eclipsing Detached Double White Dwarf Binary NLTT~11748}

\slugcomment{Accepted for publication in ApJ Letters, 2010 May 11}

%\author{Justin D. R. Steinfadt}
%\affil{Department of Physics, Broida Hall,\\University of California,
%Santa Barbara, CA 93106;\\jdrs@physics.ucsb.edu}

%\author{Lars Bildsten}
%\affil{Kavli Institute for Theoretical Physics and Department of Physics, Kohn Hall,\\University of California, Santa Barbara, CA %93106;\\bildsten@kitp.ucsb.edu}

%\author{Steve B. Howell}
%\affil{WIYN Observatory and National Optical Astronomy Observatory,\\950 N. Cherry Ave., Tucson,  AZ 85719;\\howell@noao.edu}

\author{Justin D. R. Steinfadt\altaffilmark{1}}

\author{David L. Kaplan\altaffilmark{2,3}}

\author{Avi Shporer\altaffilmark{1,4}}

\author{Lars Bildsten\altaffilmark{1,2}}

\author{Steve B. Howell\altaffilmark{5}}

\altaffiltext{1}{Department of Physics, Broida Hall,\\University of California, Santa Barbara,  CA 93106, USA}
\altaffiltext{2}{Kavli Institute for Theoretical Physics, Kohn Hall,\\University of California, Santa Barbara,  CA 93106, USA}
\altaffiltext{3}{Hubble Fellow}
\altaffiltext{4}{Las Cumbres Observatory Global Telescope Network,\\6740 Cortona Drive Suite 102, Santa Barbara,  CA 93117, USA}
\altaffiltext{5}{National Optical Astronomy Observatory,\\950 North Cherry Avenue, Tucson,  AZ 85719, USA}

% -----------------------------------------------------------
% -----------------------------------------------------------

\begin{abstract}

We report the discovery of the first eclipsing detached double white dwarf (WD) binary. In a pulsation search, the low-mass helium core WD NLTT~11748 was targeted for fast ($\approx 1$ minute) differential photometry with the Las Cumbres Observatory's Faulkes Telescope North. Rather than pulsations, we discovered $\approx 180$ s 3\%--6\% dips in the photometry. Subsequent radial velocity measurements of the primary white dwarf from the Keck telescope found variations with a semi-amplitude $K_1=271\pm 3 \, \kms$, and confirmed the dips as eclipses caused by an orbiting WD with a mass $M_2=0.648$--$0.771 \, M_\odot$ for $M_1=0.1$--$0.2 \, M_\odot$. We detect both the primary and secondary eclipses during the $P_{\rm orb}=5.64$\,hr orbit and measure the secondary's brightness to be $3.5\%\pm 0.3 $\% of the primary at SDSS-$g^\prime$. Assuming that the secondary follows the mass--radius relation of a cold C/O WD and including the effects of microlensing in the binary, the primary eclipse yields a primary radius of $R_1=0.043$--$0.039 \, R_\odot$ for $M_1=0.1$--$0.2 \, M_\odot$, consistent with the theoretically expected values for a helium core WD with a thick, stably burning hydrogen envelope.  Though nearby (at $\approx150$\,pc), the gravitational wave strain from NLTT~11748 is likely not adequate for direct detection by the {\it Laser  Interferometer Space Antenna}. Future observational efforts will determine $M_1$, yielding accurate WD mass--radius measurement of both components, as well as a clearer indication of the binary's fate once contact is reached.
\end{abstract}

%\keywords{binaries: eclipsing--- stars: white dwarf--- stars: individual: NLTT~11748}

% -----------------------------------------------------------
% -----------------------------------------------------------

%\renewcommand{\baselinestretch}{1}

\section{Introduction}

Double white dwarfs (WDs) in tight enough ($P_{\rm orb}< {\rm day}$) binaries to reach contact in a Hubble time are expected on theoretical grounds \citep{nel01}, and are presumed to be the progenitors of highly variable objects: R~CrB stars, AM~CVn binaries, and Type Ia supernovae \citep{ibe84,web84}. However, examples of these systems are rare, with only 10 known prior to our work \citep{nel01,bad09,mul09,kil09,mar10,kul10}. Many of these binaries are of immediate interest for the {\it Laser Interferometer Space Antenna} ({\it LISA}), providing `verification' sources loud enough to be detected in space \citep{nel09}.

Spectral measurements of the high proper motion object \object[NLTT 11748]{NLTT~11748} by \citet{kaw09} revealed it to be a low-mass ($<0.2 \, M_\odot$) helium core WD with $\log g=6.2 \pm 0.15$ and $T_{\rm eff}=8540\pm 50$\,K and an H-rich (DA) atmosphere. The strong expectation that low-mass He WDs can only be formed in common envelope events led \citet{kaw09} to note that this object was likely a binary. Indeed, they found a $40\, \kms$ velocity difference between two measured spectra.   NLTT~11748 was originally targeted in a ZZ~Ceti like pulsation search based on theoretical calculations and the $\log g$ and $T_{\rm eff}$ measurements \citep{ste10}. In this failed pulsation search, our rapid  ($\approx 1$ minute) differential photometry with the Las Cumbres Observatory's Faulkes Telescope North (FTN) revealed $\approx 180$ s 3\%--6\% dips, which were confirmed by Keck spectroscopy to be primary and secondary eclipses from a $\approx 0.7 \, M_\odot$ faint C/O WD companion orbiting at 5.6\,hr. 

NLTT~11748 joins the growing class \citep{kaw06,kil07,mul09,kil09,mar10,kul10,bad09} of WD binaries where the low-mass WD is a very low-mass ($<0.2 \, M_\odot$) helium core WD. These are made when the lower mass star overfills its Roche lobe at the base of the red giant branch \citep{van96,cal98,bas06a}, triggering a common envelope event that leads to an in-spiral of the now exposed low-mass He core to a shorter orbital period. Theoretical work has found a dichotomy in the evolution of these low-mass He core WDs \citep{dri99,ser02,pan07} that dramatically impact their cooling rate. For masses $\gtrsim$0.2\,$M_{\odot}$ (metallicity dependent), the H envelope undergoes shell flashes that reduce its mass, eventually allowing the WD to cool rapidly. However, less massive ($<0.2 \, M_\odot$) WDs undergo stable H burning for gigayears, dramatically slowing their evolution and keeping them brighter for much longer than expected.

Our observations are summarized in Section \ref{sec:obs}, demonstrating that the companion to NLTT~11748 is a cold ($T_{\rm eff} \lesssim 7400$\,K), old (1.5--3\,Gyr) C/O WD. We show in Section \ref{sec:res} that eclipse modeling yields the first radius measurement of an extremely low-mass WD.  We close in Section \ref{sec:conc} by discussing the value of additional measurements for constraining WD mass--radius relations. 

% FIGURE ONE
\begin{figure}
	\centering
	\epsscale{1.0}
	\plotone{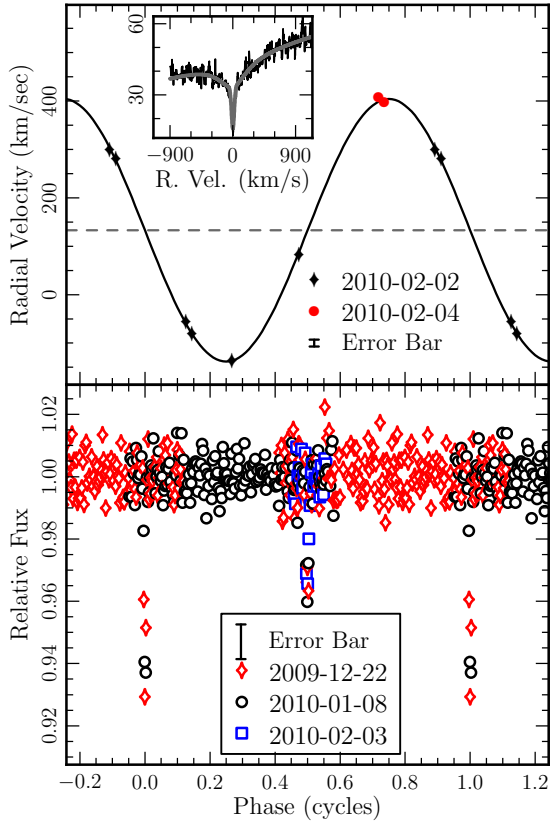}
	\caption{Top panel: phased radial velocities from Keck/HIRES H$\alpha$ measurements over two consecutive nights.  The black line is the ($e=0$) sinusoid with $K_1=271 \, \kms$ amplitude and the $v_r = 133 \, \kms$ systemic velocity offset.  Inset shows the phase combined H$\alpha$ spectrum, in arbitrary flux units, with fitting function over plotted.  Bottom panel: phased photometric light curve for all three nights of data.  The error bar represents $4-5$\,mmag uncertainty.}
	\label{fig:rvlc}
\end{figure}

\begin{deluxetable}{l c c c}
\tablecaption{Measured and Derived Quantities for NLTT~11748\label{tbl:params}}
\tablehead{
\colhead{Quantity} & \mc{3}{c}{Value}\\
\tableline
\mc{4}{c}{HIRES spectra}
}
\startdata
Rad.\ vel.\ amp., $K_1$ (km\,s$^{-1}$)  & \mc{3}{c}{271(3)} \\
Sys. radial velocity, $v_{r}$ (km\,s$^{-1}$)  & \mc{3}{c}{133(2)} \\
$\chi^2$/dof         & \mc{3}{c}{4.5/6} \\
\cutinhead{FTN photometry}
Time of prim.\ ecl. (BJD TDB)  & \mc{3}{c}{2,455,196.87828(7)} \\
Period (days)  & \mc{3}{c}{0.2350606(11)} \\
Ephemeris $\chi^2$/dof         & \mc{3}{c}{3.5/3} \\
Primary eclipse depth, $d_1$  & \mc{3}{c}{0.067(3)} \\
Secondary eclipse depth, $d_2$  & \mc{3}{c}{0.034(2)} \\
Primary ecl.\ duration, $\tau_1$ (s) & \mc{3}{c}{180(6)\phz} \\
Secondary ecl.\ duration, $\tau_2$ (s) & \mc{3}{c}{185(10)} \\
$F_2/F_1$ (in SDSS-$g^\prime$)  & \mc{3}{c}{0.035(3)} \\
Out of eclipse\tablenotemark{a} $\chi^2$/dof         & \mc{3}{c}{404.8/372} \\
\cutinhead{Combined data (assuming Mass of Primary $M_1=0.15M_{\odot}$)}
Limb darkening coefficient, $u_{\rm LD}$  &  0.0  &  0.3  &  0.5  \\
Mass of secondary, $M_2$ ($M_{\odot}$)   & 0.71(2)  &  0.71(2)  &  0.71(2)\\
Inclination (deg)  & 89.90(11) & 89.88(11) & 89.87(11)\\
Radius of primary, $R_1$ ($R_{\odot}$)   & 0.0393(9) & 0.0406(9) & 0.0415(9)\\
%Radius of Secondary $R_2$\tablenotemark{b} ($R_{\odot}$) \dotfill & 0.0108(2) & 0.0108(2) & 0.0108(2) \\
$\chi^2$/dof\tablenotemark{b}         & 285.1/227 & 279.5/227 & 276.5/227\\
Distance\tablenotemark{c} (pc)  & \mc{3}{c}{150(32)} \\
Sys.\ kin.\ $(U,\,V,\,W)$\tablenotemark{d} (km\,s$^{-1}$)  & \mc{3}{c}{(-151(9), -183(41), -34(5))} \\
\enddata
\tablecomments{Quantities in parentheses are 1$\sigma$ uncertainties on the last digit.}
\tablenotetext{a}{Excluding $\pm0.005$ cycles around each eclipse.}
%\tablenotetext{b}{Derived from $M_2$ using a CO WD mass-radius relation.}
\tablenotetext{b}{Result of fitting phases $\pm 0.1$ cycles around each eclipse.}
\tablenotetext{c}{Scaled from \citet{kaw09} using our $R_1$.}
\tablenotetext{d}{Calculated from the proper motion \citep{kaw09} corrected to the local standard of rest \citep{hog05} and our updated distance.}
\end{deluxetable}

% -----------------------------------------------------------
% -----------------------------------------------------------

\section{Observations}\label{sec:obs}

We targeted NLTT~11748 as part of an observational search for ZZ Ceti like pulsations from low-mass He core WDs \citep{ste08b,ste10}.  Our photometric discovery led to our Keck spectroscopy, both described here.

\subsection{Faulkes Telescope North Photometry}

Photometric observations were done with the 2 m robotic (FTN) \citep{lew10}, part of the Las Cumbres Observatory Global Telescope (LCOGT) network\footnote{http://lcogt.net}, on Haleakala, Hawaii. We used the Merope camera with 2$\times$2 pixel binning for a pixel scale of $0\farcs28 \, {\rm pixel}^{-1}$ and a $4.75^\prime \times 4.75 ^\prime$ field-of-view. Our observational setup, consisting of the SDSS-$g^\prime$ filter and 45\,s exposure time (with $\approx$22\,s dead time), was aimed at detecting pulsations, expected to be at the few minutes timescale \citep{ste08b,ste10}.  Preliminary reduction, including bias and flat-field corrections using standard IRAF\footnote{IRAF is distributed by the National Optical Astronomy Observatory, which is operated by the Association of Universities for Research Astronomy, Inc., under cooperative agreement with the National Science Foundation.  http://iraf.noao.edu} routines, is done automatically at night's end. 

Our first observation on 2009 December 22 (night 1) was for 4\,hr.  We immediately identified two dips in the light curve, $\approx$3\,minutes in duration and a few percent in depth, indicative of eclipses. To verify the existence of eclipses we observed the system again on 2010 January 8 (night 2), for 3.5\,hr, revealing two similar eclipses at the expected times based on the previous observations. An additional 30-minute observation was obtained on 2010 February 3 (night 3).

Photometric processing was carried out using IRAF aperture photometry and the VAPHOT package of routines \citep{dee01}.  Three comparison stars were selected to yield a light curve with the smallest variance.  In 45\,s exposures, each of these stars accumulated $\approx4$$\times$10$^5 \, e^-$ while NLTT~11748 accumulated $\approx7$$\times$10$^5 \, e^-$.  The differential light curve for NLTT~11748 was constructed using the scheme detailed in Steinfadt et al. (2008b, inspired by \citealt{sok01}).  A de-trending second-order polynomial removed uncorrected long-term variations, such as color--airmass effects.  For all data points outside of eclipses we find a $\chi^2/$dof of 404.8/372, indicating a reasonable estimate of our uncertainties.

%After considering the radial velocity measurements in Section \ref{sec:obsrv}, we found that Nights 1 and 2 had primary and secondary eclipses, while Night 3 shows a secondary eclipse.  Using the technique of \citet{kwe56} we fit the transit centers of each eclipse, and combined all eclipse timings (Primary (UTC~HJD): 2,455,188.885429(89), 2,455,205.80992(29); Secondary (UTC~HJD): 2,455,188.76805(16), 2,455,205.92705(19), 2,455,231.78410(19)) in a linear least-squares analysis to find the orbital period of $P_{\rm orb}=5.641454 \pm 0.000026$\,hrs.  Fitting ephemerides using the primary and secondary eclipses separately give consistent results with our reported ephemeris (Table \ref{tbl:params}).  The resulting phased light curve from all three nights of data is shown in Figure \ref{fig:rvlc}.

After considering the radial velocity measurements in Section \ref{sec:obsrv}, we found that nights 1 and 2 had primary and secondary eclipses, while night 3 shows a secondary eclipse.  Using the technique of \citet{kwe56} we fit the transit centers of each eclipse, and combined all eclipse timings (primary (BJD TDB): 2,455,188.886207(89), 2,455,205.81071(29); secondary (BJD TDB): 2,455,188.76882(16), 2,455,205.92783(19), 2,455,231.784879(19)) in a linear least-squares analysis to find the orbital period of $P_{\rm  orb}=5.641454 \pm 0.000026$\,hr (barycentering used the JPL DE405 ephemeris).  Fitting ephemerides using the primary and secondary eclipses separately gives consistent results with our reported ephemeris (Table \ref{tbl:params}).  The resulting phased light curve from all three nights of data is shown in Figure \ref{fig:rvlc}.

% -----------------------------------------------------------
% -----------------------------------------------------------

\subsection{Keck Spectroscopy}\label{sec:obsrv}

We observed NLTT~11748 with High Resolution Echelle Spectrometer (HIRES, \citealt{vog94}) on the 10-m Keck~I telescope 6 times on 2010 February 2 and twice on 2010 February 4.  All observations were taken with the same grating tilt, integrating for 5\,minutes, binning by 3 pixels in the spatial direction, and with a $0\farcs86\times14\arcsec$ slit (giving $R\approx 50,000$).  The wavelength solution was based on a Th--Ar lamp, accurate to $<0.1$\,pixel (with $1.4\,\kms\,{\rm pixel}^{-1}$), and the data covered $3600-8000\,$\AA.  The stability of the solution was monitored to high precision by G.~Marcy (private communication, 2010).

We used the H$\alpha$ line for our radial velocity measurements, although other lines were detected at lower significance.  We initially identified the velocity from each spectrum by eye, then shifted each spectrum to zero velocity and combined them.  We fit the combined H$\alpha$ spectrum with the sum of a broad ($\approx 520\,\kms$ FWHM) Lorentzian and a narrow ($\approx 47\,\kms$ FWHM) Gaussian with roughly equal depths.  Using this template, we then measured radial velocities and fit them to a sinusoid with the period and phase constrained by the eclipse timing.  We then iterated this procedure, using the velocities from the radial-velocity fit to construct the template, fitting the template shape, fitting the individual velocities, and fitting the radial velocity curve.  This converged quickly: after one iteration all changes were $<1\,\kms$. 

We obtain a $\chi^2$ for the radial velocity fit (assuming a circular orbit and an accurate ephemeris from the eclipse photometry) of 4.5 for 6 dof.  Our final radial-velocity amplitude was $K_1=271\pm3\,\kms$ yielding a mass function $0.48(2) \, M_{\odot}$, or an $M_2=0.71(2) \, M_{\odot}$ secondary for a fiducial $M_1 = 0.15 \, M_{\odot}$ primary.  We computed the systemic radial velocity from the data, correcting  to the solar system barycenter,  finding $v_r=133\pm2\,\kms$.  We do not correct for gravitational redshift as uncertainties in $M_1$ and $R_1$ are too high and we expect a shift of only $\approx 2 \, \kms$.  We checked our velocities by fitting only the broad component of the H$\alpha$ line or the H$\gamma$ line, and both gave consistent results although with a factor of 6 lower precision.  We limit the eccentricity of the orbit to $<0.06$ at 3$\sigma$.

At the primary's orbital velocity, the width of the narrow H$\alpha$ component may be broadened by orbital motion: the maximum derivative is $2\pi K_1/P_{\rm orb}=0.08\,\kms\,{\rm s}^{-1}$. Compared with the narrow H$\alpha$  component, $\approx 47 \kms$ FWHM, this effect is unlikely to affect our data (5-minute exposures) but might have slightly influenced the $\log g$ measurement of \citet{kaw09}.

% -----------------------------------------------------------
% -----------------------------------------------------------

\section{Binary Parameter Analysis and Results}\label{sec:res}
\label{sec:binpa}

% FIGURE TWO
\begin{figure}
	\centering
	\epsscale{1.0}
	\plotone{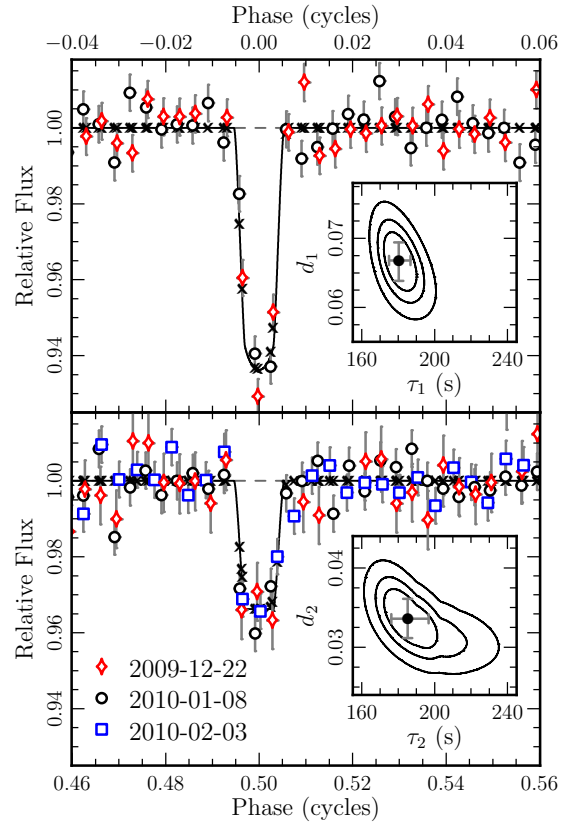}
	\caption{Phased light curves of the primary (top panel) and secondary (bottom panel) eclipses.  The widths of each data point represent the exposure duration.  The black line is the light curve model for the $u_{\rm LD}=0.3$ parameters in Table \ref{tbl:params}.  The crosses show the average of the model over the exposure duration of that associated data point.  It is the differences between these points and those measurements that are considered when calculating $\chi^2$.  Insets display the 1$\sigma$, 2$\sigma$, and 3$\sigma$ contours of the $\chi^2$ minimization for the eclipse depth and width parameterized light curve fitting.  The error bars are the 1$\sigma$ error bars on the individual parameters. }
	\label{fig:ecllc}
\end{figure}

The deepest (primary) eclipse is the transit of the primary object (1) by the secondary (2), and the shallowest (secondary) is the occultation of the secondary by the primary.  There are six relevant binary parameters to be fitted: the primary and secondary radii and masses ($R_1$, $R_2$, $M_1$ and $M_2$); flux ratio (for our filter/CCD response) $F_2/F_1$; and inclination $i$.  We consider the orbital period, $P_{\rm orb}$, and the linear limb darkening coefficient, $u_{\rm LD}$, as fixed.  We presume $u_{\rm LD}=0.3$ but will show how its uncertainty ($u_{\rm LD} \approx 0.2-0.6$, \citealt{lit06,lit07}, and \citealt{max07}) modifies our desired parameters.  The eclipse light curve in Figure \ref{fig:ecllc} shows the secondary eclipse as flat-bottomed while the primary eclipse shows the distinct signature of limb darkening.  This indicates that the eclipses are likely total (confirmed by our inclination measurements) and $R_2 < R_1$, $F_2/F_1 \ll 1$, and $i \approx 90^{\circ}$.

We initially fit the eclipses with a parameterization of the linear limb darkening law \citep{van93} for the primary eclipse and a box function for the total secondary eclipse, yielding the depths, $d_1$ and $d_2$, and durations, $\tau_1$ and $\tau_2$.  The 1$\sigma$ results are reported in Table \ref{tbl:params} with the 1$\sigma$, 2$\sigma$, and 3$\sigma$ contours plotted in the insets of Figure \ref{fig:ecllc}.  This allows us to directly measure $F_2/F_1$, as $1-d_2 = F_1/\left( F_1 + F_2 \right)$.

The sparsity of our data in eclipse requires additional assumptions about the properties of the secondary.  The flat bottom secondary eclipse unambiguously gives a low flux ratio (at SDSS-$g^\prime$) of $F_2/F_1=0.035(3)$ which implies that the value of the primary eclipse total depth must be dominated by the radii ratio.  Our measured primary eclipse depth implies $R_2/R_1 \approx 1/4$, further corroborated by the unseen (and thereby fast) ingress and egress in the secondary eclipse.  Our measurements of $F_2/F_1$ and $R_2/R_1$ constrain the temperature of the secondary relative to the 8500\,K primary.  Taking SDSS-$g^{\prime}$ measurements as bolometric ($L_2/L_1=F_2/F_1$) gives $T_2\leq7400\,$K, which we confirm by using the synthetic photometry of \citet{hol06}, constraining the flux ratio of model atmospheres integrated over the filter passband.  The \citet{kaw09} spectrum constrains the primary to be a low-mass, $\lesssim 0.25 \, M_{\odot}$, likely He WD with a radius $\approx 0.04 \, R_{\odot}$.  This implies an $R_2 \approx 0.01 \, R_{\odot}$ companion.  Therefore, the secondary cannot be a main sequence (MS) star as \citet{kaw09} constrain such an MS star to have $M<0.2 \, M_{\odot}$ based on Two Micron All Sky Survey photometry, breaking our mass function and radius constraints.  A brown dwarf would break our mass function and radius constraints.  A neutron star is far too small and its luminosity contribution insufficient.  Therefore, a cold C/O WD is the only solution, meeting the radius, mass, and luminosity constraints.

% FIGURE THREE
\begin{figure}
	\centering
	\epsscale{1.00}
	\plotone{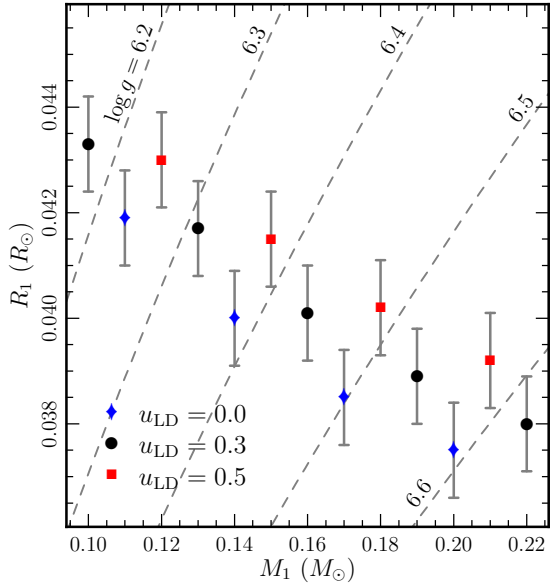}
	\caption{Primary radius vs. primary mass.  The uncertainties are 1$\sigma$ uncertainties assuming the values and uncertainties of $K_1$ and $F_2/F_1$ given in Table \ref{tbl:params} and assuming a secondary radius derived from its mass.  Dashed lines are constant $\log g$.}
	\label{fig:rpvsmp}
\end{figure}

We powerfully constrain the binary parameters by directly integrating the faces of the primary and secondary and model our light curve for any arbitrary inclination and phase by assuming a linear limb
darkening law.  We do this with a code originating from \citet{ste08a} and \citet{man02}, including the effects of microlensing in the binary (following \citealt{marsh01} and \citealt{agol02}).  For the approximate parameters above, the Einstein radius is $R_{\rm  E}=\sqrt{4 G M_2 a/c^2}\approx 0.003\,R_\odot$, giving $R_1/R_{\rm  E}\approx 13$ and $R_2/R_{\rm E}\approx 3$, leading to a 1\% magnification of the primary at conjunction.  Our measurement of $K_1$ and $F_2/F_1$ reduces our six-parameter system to four-parameter system.  Our surface temperature measurement yields an age of 1.5--3.0 Gyr \citep{cha00} for the secondary allowing us to use M-R relations of cold WDs \citep{alt98} for our current work.  Though \citet{kaw09} derived an $M_1 = 0.167 \pm 0.005 \, M_{\odot}$ from their low-resolution spectrum, radial velocity broadening and theoretical model uncertainties may render this determination unreliable.  Therefore, we calculate separate $\chi^2$ grids of $R_1$ and inclination for a range of $M_1$.  Minimization of $\chi^2$ on each grid constrains $R_1$ and inclination directly and $\Delta \chi^2$ techniques yield their uncertainties.  We scale these uncertainties to consider the additional uncertainties of $F_2/F_1$ and $K_1$ using a Monte Carlo method.

The results of our direct eclipse fitting method are displayed in Figure \ref{fig:rpvsmp} with the parameters and uncertainties reported in Table \ref{tbl:params} for $M_1 = 0.15 \, M_{\odot}$.  Without including lensing the fits would be very similar, but the inclination would be slightly lower (closer to $89.4\degr$) and the radius of the primary would be slightly larger ($0.0437\,R_{\odot}$ for $u_{\rm LD}=0.3$). The $\chi^2$ minima for all $M_1$ within each $u_{\rm LD}$ data set in Figure \ref{fig:rpvsmp} are approximately consistent and as such yield no constraint on $M_1$ over this range.  Our range of $u_{\rm LD}=0.0$, 0.3, and 0.5 results in a $\approx5$\% spread in $R_1$ measurements, consistent with the $\log g= 6.2 \pm 0.15$ of \citet{kaw09}.  The $\chi^2$ value is reduced slightly as $u_{\rm LD}$ increases; however, we cannot constrain $u_{\rm LD}$ due to the sparsity of data at ingress/egress.  This will be constrained in future higher cadence data sets.  For $M_1 > 0.2 \, M_{\odot}$ the minimum $\chi^2$ values significantly increase and the quality of the eclipse fits degrade.  Therefore, our combined radial velocity and light curve data sets provide an upper limit, assuming that the secondary is a C/O WD, of $M_1 \lesssim 0.2 \, M_{\odot}$.

% -----------------------------------------------------------
% -----------------------------------------------------------

\section{Conclusions}\label{sec:conc}

Our discovery of the first eclipsing detached double WD binary has substantially constrained low-mass He WD models with radius measurements that agree with current models \citep{ser02,pan07}. Additional measurements will reveal even more about the faint $\approx 0.7 \, M_{\odot}$ C/O WD secondary. With our discovery, there are 11 confirmed double WD binaries that will merge within a Hubble time. From spectroscopic masses alone, six of these contain low-mass ($<0.25M_\odot$) He core primaries with large ($>0.03 \, R_\odot$) radii indicative of a stable H burning shell \citep{mul09,kil09,bad09,mar10,kul10}. Other than its inclination, NLTT~11748 is not extraordinary in any way relative to these, many of which offer larger solid angles for eclipse observations than NLTT~11748.  

Our analysis of this system is far from complete, but our existing constraints, Table \ref{tbl:params} and Figure \ref{fig:rpvsmp}, highlight the value of further studies. High cadence photometry, resolving the ingress/egress of each eclipse, would yield the component radii as functions of inclination and orbital semimajor axis. If future spectroscopic observations can reveal the secondary's line features and measure its radial velocity, this would allow precise measurements of the masses of both components.  Detection of eclipses in different filters, both bluer and into the infrared, would additionally constrain the temperature of the secondary further as well as its age.  

Model-independent mass and radius determinations for He and C/O WDs are rare.  For He WDs, there exist several pulsar binary systems for which Shapiro time delay measurements have yielded accurate mass determinations: \object[PSR B1855+09]{PSR~B1855+09} \citep{kas94}, \object[PSR J0437-4715]{PSR~J0437$-$4715} \citep{van01}, \object[PSR J0751+1807]{PSR~J0751+1807} \citep{bas06b}, \object[PSR J1012+5307]{PSR~J1012+5307} \citep{cal98,van96}, \object[PSR J1911-5958A]{PSR~J1911$-$5958A} \citep{bas06a}, and \object[PSR J1909-3744]{PSR~J1909$-$3744} \citep{jac03,jac05}. Of these, only four have radii measurements, PSR~J0437$-$4715, PSR~J0751+1807, PSR~J1911$-$5958A and PSR~J1909$-$3744.  However, all rely upon either distance measurements and evolutionary model-dependent luminosities or gravity measurements derived from atmospheric models.  Therefore, He WD evolutionary models \citep{alt98,ser02,pan07} remain largely unconstrained. For C/O WDs, mass and radius measurements have been more frequent \citep{sch96,pro98,pro02,cas09} and all rely on distance measurements via parallax or  cluster membership, surface brightness measurements and detailed knowledge of each spectrum.  To reach the precisions required to constrain theoretical models, these measurements must rely upon atmospheric model spectra.  Only recently \citet{par10} have derived precise measurements of mass and radius in a model-independent way in the eclipsing binary system \object[NN Ser]{NN~Ser}.  NLTT~11748 offers an additional C/O WD system for model-independent mass and radius measurements as well as the first such system for He WDs.

Gravitational wave emission will bring this system into contact in $13.8-6.3$\,Gyr (for $M_1=0.1-0.2 \, M_\odot$).  With a mass ratio of $0.15/0.71=0.2$, stable mass transfer at contact is likely to occur and create an AM~CVn binary \citep{nel01,mar04}.  The known orbital period, inclination, and distance ($d\approx150$\,pc) yield the gravitational wave strain at Earth, $h=(3.6-7.9)\times 10^{-23}$ \citep{tim06,roe07} at a frequency $\nu\approx 10^{-4}$\,Hz for $M_1=0.1-0.2 \, M_\odot$. Though not `louder' than the verification sources tabulated by \citet{nel09}, NLTT~11748's accurate ephemeris allows for a coherent folding of the {\it LISA} data over the mission duration.

\acknowledgments
 
We thank the referee for helpful comments and suggestions.  This paper uses observations obtained with facilities of the Las Cumbres Observatory Global Telescope.  Some of the data presented herein were obtained at the W.M. Keck Observatory, which is operated as a scientific partnership among the California Institute of Technology, the University of California, and NASA. The Observatory was made possible by the generous financial support of the W.M. Keck Foundation.  We thank Geoff Marcy and Howard Isaacson for allowing us to use their Keck time for our spectroscopy.  We thank Mansi Kasliwal, Shri Kulkarni, Tom Marsh, and Chris Watson for their gracious offers of telescope time.  We thank E.~Agol for helpful discussions and providing lensing/occultation software.  D.L.K. was supported by NASA through Hubble Fellowship grant \#51230.01-A awarded by the STScI, which is operated by AURA, for NASA, under contract NAS 5-26555. This work was supported by the National Science Foundation under grants PHY 05-51164 and AST 07-07633.

{\it Facilities:} \facility{Keck:I (HIRES)}, \facility{FTN (Merope)}

% -----------------------------------------------------------
% -----------------------------------------------------------

\end{document}